\documentclass[a4paper,12pt]{article}

\usepackage[utf8]{inputenc}
\usepackage[T1]{fontenc}
\usepackage[english]{babel}

\usepackage{indentfirst}
\usepackage{amsmath}
\usepackage{amssymb}
\usepackage{amsthm}
\usepackage{graphicx}
\usepackage[font=small]{caption}
\usepackage{subcaption}
\usepackage{psfrag}{}
\usepackage{color}
\usepackage{pgf}

\allowhyphens

\newcommand{\valos}{\mathbb{R}}

\newcommand{\ordo}{\mathcal{O}}

\newcommand{\fa}{\mathfrak{a}}

\newcommand{\ket}[1]{{\left|#1\right\rangle}}
\newcommand{\bra}[1]{{\left\langle #1\right|}}

\newcommand{\vev}[1]{\left\langle #1 \right\rangle}

\newtheorem{conj}{Conjecture}


\makeatother

\begin{document}

\numberwithin{equation}{section}

\author{M\'arton Mesty\'an$^1$, Bal\'azs Pozsgay$^1$\\
~\\
 $^{1}$MTA--BME \textquotedbl{}Momentum\textquotedbl{} Statistical
Field Theory Research Group\\
1111 Budapest, Budafoki út 8, Hungary
}
\title{Short distance correlators in the XXZ spin chain for arbitrary
  string distributions}

\maketitle

\abstract{
In this letter we consider expectation values of local correlators in highly
excited states of the spin-1/2 XXZ chain. Assuming that the string
hypothesis holds we formulate the following conjecture: The
correlation functions can be computed using the known factorized formulas
of the finite temperature situation, if the building blocks are computed
via certain linear integral equations using the string densities
only. 
We prove this statement for the nearest neighbour z-z 
correlator for states with arbitrary string densities. Also, we check
the conjecture numerically for other correlators in the finite
temperature case. Our results pave the way towards the computation of  the stationary
values of correlators in non-equilibrium situations using the 
quench action approach.
}

\section{Introduction}

The analytical calculation of correlation functions of the spin-1/2 XXZ chain 
is
by now a very advanced subject. There
are multiple integral formulas available for mean values of local operators
for both the ground state \cite{Jimbo-Miwa-book,Maillet-xxz-ground-state-corr-review}
and in the finite temperature case \cite{QTM1}. Moreover it is known
that these multiple integrals can be factorized, i.e. they can be
expressed as sums of products of simple integrals (see
\cite{XXZ-factorization-recent-osszefoglalo} and references
therein). This provides a very efficient way for the numerical
evaluation of the local correlations.

Previous research on this subject only considered the ground state or finite temperature
mean values. However, recent advances in the study of non-equilibrium
processes 
provide motivation to determine the correlations in more general
situations, for example in the case of quantum quenches \cite{quench-action}. Ideally one
would like to obtain exact results for the time-dependent correlators
after a quench, but this is beyond the reach of the present
methods. A simpler task is derive the mean values in the long-time
limit. 

The Generalized Gibbs Ensemble (GGE) hypothesis \cite{rigol-gge}
states that  the stationary values of local correlators can be
evaluated using a modified statistical ensemble which incorporates all
higher local charges of the model with appropriate
Lagrange multipliers. 
The papers \cite{sajat-xxz-gge,essler-xxz-gge,fagotti-collura-essler-calabrese}
developed generalizations of the  Quantum Transfer Matrix
 method (originally devised for the finite temperature 
problem \cite{kluemper-review}) to compute local correlators in the GGE for quantum quenches
 in the XXZ chain. While \cite{sajat-xxz-gge} and \cite{essler-xxz-gge} employed
 approximations, the authors of
 \cite{fagotti-collura-essler-calabrese} were able to calculate the
 exact predictions of the GGE in a number of different cases.
It is important to note that all of these papers used the factorized formulas for the correlators
found in \cite{XXZ-finite-T-factorization}.

In \cite{quench-action} a different approach was suggested which is
based on first principles. If the overlaps between the initial state
and the eigenstates of the final Hamiltonian are known, then the
analysis of the so-called quench action (overlaps and entropy
combined) provides the Bethe root
distributions of sample states which determine the time evolution at large times. Such
a calculation is completely analogous to the minimalization of the
free energy using the Thermodynamic Bethe
Ansatz (TBA), but the resulting root distributions are typically very
different from the thermal case 
\cite{caux-stb-LL-BEC-quench}. 

Concerning the XXZ chain one drawback of the quench action approach is
that it only provides the string densities, but up to now it has not
been known how
to compute the correlation
functions which are the actual measurable quantities. Here we make an
attempt to fill this gap. Our work is partly motivated by the papers
\cite{Caux-Neel-overlap1,Caux-Neel-overlap2} where 
simple determinant formulas have been found for the
overlaps between the N\'eel state and the Bethe states, thus making
the determination of the saddle point distributions possible \cite{caux-talk,JS-oTBA,sajat-oTBA}.

The paper is organized as follows.
In Section 2 we introduce the model and its coordinate Bethe Ansatz
solution. In Section 3 we compute an integral formula for the nearest
neighbour z-z correlator, which holds for an arbitrary distribution of
Bethe roots. The finite temperature situation is considered in Section
4, where we present the QTM formulas for the local correlators and
the TBA result for $\vev{\sigma_1^z\sigma_2^z}_T$. In Section 5 we introduce a
generalization of the formulas of Section 2 and present two
Conjectures for the local correlators. Finally, Section 6 includes the
discussion of our results. Numerical data for the correlators in
the finite temperature case are presented in the Appendix.

\section{Eigenstates and root densities}

Consider the XXZ spin-$1/2$ chain with Hamiltonian
\begin{equation}
  \label{H}
  H_{XXZ}=\sum_{j=1}^{L} \left\{
\sigma_j^x\sigma_{j+1}^{x}+\sigma_j^y\sigma_{j+1}^{y}+\Delta
(\sigma_j^z\sigma_{j+1}^{z}-1)
\right\}+h\sum_{j=1}^L (\sigma_j^z-1).
\end{equation}
Here $h$ is a longitudinal magnetic field and $\Delta=\cosh(\eta)$ is
the anisotropy parameter. For simplicity 
we only consider the regime with $\Delta>1$ where $\eta\in\valos^+$. In
this work periodic boundary conditions are assumed.

Eigenstates can be constructed by the different forms of the Bethe
Ansatz \cite{KorepinBook}. 
Choosing the vector 
$\ket{F}=\ket{++\dots  +}$
as a reference state an eigenstate with $N$ down spins 
  can be characterised by a set of rapidities
(quasi-momenta) $\{\lambda_1,\dots,\lambda_N\}$ which describe the
propagation of the interacting spin waves. In coordinate Bethe Ansatz
the explicit wave function can be written as 
\begin{equation}
  \label{BA}
\Psi_{N}(\lambda_1,\dots,\lambda_N|s_1,\dots,s_N)=\sum_{P\in\sigma_N} 
\prod_j 
\left(\frac{\sin(\lambda_{P_j}+i\eta/2)}{\sin(\lambda_{P_j}-i\eta/2)}\right)^{s_j}
\prod_{j>k} \frac{\sin(\lambda_{P_j}-\lambda_{P_k}-i\eta)}{\sin(\lambda_{P_j}-\lambda_{P_k})}.
\end{equation}
Here $s_j$ denote the positions of the down spins, and we assume
$s_j<s_k$ for $j<k$. Periodic boundary conditions impose the Bethe
equations:
\begin{equation}
  \label{BAe}
P(\lambda_j)^L
\prod_{k\ne j}
S(\lambda_j-\lambda_k)=1,
\end{equation}
where
\begin{equation*}
  P(\lambda)=e^{ip(\lambda)}=\frac{\sin(\lambda+i\eta/2)}{\sin(\lambda-i\eta/2)}
\end{equation*}
is the propagator of a single down spin and the function
\begin{equation*}
S(\lambda)=\frac{\sin(\lambda-i\eta)}{\sin(\lambda+i\eta)}
\end{equation*}
describes the scattering of the spin waves.

If the Bethe equations hold then the energy eigenvalue is given by
\begin{equation}
\label{BAee}
  E_\Psi=-2Nh+\sum_j e(\lambda_j),\quad\text{where}\qquad
e(u)=\frac{4\sinh^2\eta}{\cos(2u)-\cosh\eta}.
\end{equation}

The string hypothesis \cite{Takahashi-book} states that in a large volume typical
eigenstates consists of spin waves with real rapidities $\lambda_j$ and/or bound
states of $n$ spin waves such that the rapidities of the constituents
 are
 \begin{equation*}
    \{\lambda\}_n=x-\frac{n-1}{2}i\eta+i\delta_1,x-\frac{n-3}{2}i\eta+i\delta_2,
\dots, x+\frac{n-1}{2}i\eta+i\delta_n.
  \end{equation*}
Here $x\in [-\pi/2,\pi/2]$ is called the string center and the parameters
$\delta_j$ are string deviations which are exponentially small in
$L$ and can be neglected in the thermodynamic limit.

The string hypothesis allows us to write down a set of equations for
the string centers only. Let us assume that the Bethe state consists
of $n$ ''particles'' such that the string lengths are $\alpha_k$,
$k=1\dots n$, and the string centers are given by $x_k$. The total
number of down spins is
\begin{equation*}
  N=\sum_{k=1}^n \alpha_k.
\end{equation*}
Then we have 
\begin{equation}
    \label{BAe2}
P_{\alpha_j}(x_j)^L
\prod_{k\ne j}
S_{\alpha_j\alpha_k}(x_j-x_k)=1,\qquad\qquad j=1\dots n.
\end{equation}
Here
\begin{equation*}
  P_\alpha(\lambda)=\frac{\sin(\lambda+i\alpha\eta/2)}{\sin(\lambda-i\alpha\eta/2)}
\end{equation*}
and
\begin{equation*}
  S_{nm}=
  \begin{cases}
P_{-|n-m|}(P_{-(|n-m|+2)}P_{-(|n-m|+4)}\dots
P_{-(n+m-2)})^2P_{-(n+m)}     & \text{if}\quad  m\ne n\\
(P_{-2} P_{-4}\dots P_{-(2n-2)})^2 P_{-2n}   & \text{if}\quad m=n.
  \end{cases}
\end{equation*}
The set \eqref{BAe2} is called the Bethe-Takahashi equations.

Consider a large volume $L$ with a large number of particles such that
$N/L=\ordo(1)$. 
In typical eigenstates the string centers are distributed smoothly
 for each type of string. 
Let us introduce the density of roots of j-strings $\rho_{r,j}^{(0)}$ and total
density of j-string roots and holes $\rho_{j}^{(0)}$. 
 The superscript $(0)$ is introduced for
later convenience. These functions satisfy the constraint
following from the Bethe-Takahashi equations:
\begin{equation}
\label{rhoegyenlet}
  \rho^{(0)}_k(u)=-s^{(0)}_k(u)-\sum_{l=1}^\infty \int_{-\pi/2}^{\pi/2} 
\frac{d\omega}{2\pi} \varphi_{kl}(u-\omega) \rho^{(0)}_{r,l}(\omega).
\end{equation}
Here
\begin{equation*}
    s^{(0)}_{k}(u)=(-i)\frac{\partial}{\partial u}\log P_k(u)=
(-i) \left[
\cot(u+ik\eta/2)-\cot(u-ik\eta/2)
\right]
\end{equation*}
and $\varphi_{nm}(u)=\frac{\partial }{\partial u}(-i)\log S_{nm}(u)$. Explicitly
\begin{equation}
\label{varphikeplet}
  \varphi_{nm}=
  \begin{cases}
-(s^{(0)}_{|n-m|}+2s^{(0)}_{|n-m|+2}+2s^{(0)}_{|n-m|+4}+\dots+2s^{(0)}_{n+m-2}+s^{(0)}_{n+m})    & \text{if}\quad  m\ne n\\
-(2s_2^{(0)}+2s_4^{(0)}+\dots+2s_{2n-2}^{(0)}+s_{2n}^{(0)})   & \text{if}\quad m=n.
  \end{cases}
\end{equation}
The total number of down spins is given by
\begin{equation*}
   \frac{N}{L}=
\sum_{k=1}^{\infty} k \int \frac{du}{2\pi} \rho^{(0)}_{r,k}(u).
\end{equation*}
It is useful to define the functions $\eta_k(u)$ as
\begin{equation}
\label{etak}
\frac{1}{1+ \eta_k(u)}=\frac{\rho^{(0)}_{r,k}(u)}{\rho^{(0)}_{k}(u)}
\end{equation}
such that \eqref{rhoegyenlet} can be written as
\begin{equation}
\label{rhoegyenlet2}
  \rho^{(0)}_k(u)=-s^{(0)}_k(u)-\sum_{l=1}^\infty \int_{-\pi/2}^{\pi/2} 
\frac{d\omega}{2\pi} \varphi_{kl}(u-\omega) 
\frac{\rho^{(0)}_{l}(\omega)}{1+\eta_l(\omega)}.
\end{equation}
The ratio $1/(1+\eta_k)$ can be interpreted as a ''filling fraction''
for the k-string centers.

We note that the functions $s^{(0)}_k$ are proportional to the
$j$-string energies which follows from the fact that for a single
rapidity
\begin{equation*}
  e(\lambda)=2\sinh(\eta) s^{(0)}_1(\lambda).
\end{equation*}
For further use we introduce the operator 
\begin{equation}
  \label{Q2}
Q_2=\frac{1}{2\sinh\eta} H
\end{equation}
 whose single-particle eigenfunctions coincide with $s^{(0)}_j$:
\begin{equation*}
  \vev{Q_2}_\Psi=\sum_{j=1}^N s^{(0)}_1(\lambda_j)=\sum_{k=1}^{n} s^{(0)}_{\alpha_k}(x_k).
\end{equation*}
In a large volume this relation can be written as 
\begin{equation*}
   \frac{\vev{Q_2}_\Psi}{L}=
\sum_{k=1}^{\infty} \int \frac{du}{2\pi} s^{(0)}_{k}(u)\rho^{(0)}_{r,k}(u).
\end{equation*}

To conclude this section we introduce a simplified notation for the
various convolutions and integrals which appear in this work. 
Linear integral equations of the type
\begin{equation}
\label{integral1}
  f_k(u)=g_k(u)-\sum_{l=1}^\infty \int_{-\pi/2}^{\pi/2} 
\frac{d\omega}{2\pi} \varphi_{kl}(u-\omega) 
\frac{f_k(\omega)}{1+\eta_k(\omega)}
\end{equation}
will be denoted by
\begin{equation}
\label{int2}
  f=g-
 \varphi\star \frac{f}{1+\eta}.
\end{equation}
Also, integrals of the type
\begin{equation*}
  I=\sum_{k=1}^{\infty} \int \frac{du}{2\pi} f_{k}(u)\frac{g_k(u)}{1+\eta_k(u)}
\end{equation*}
will be denoted by
\begin{equation*}
  I=f\cdot \frac{g}{1+\eta}.
\end{equation*}
With these notations the equations for the string densities
read
\begin{equation}
\label{rhoegyenlet2b}
  \rho^{(0)}=-s^{(0)}-
 \varphi\star \frac{\rho^{(0)}}{1+\eta},
\end{equation}
whereas the expectation value of $Q_2$ is expressed as
\begin{equation*}
  \frac{ \vev{Q_2}_\Psi}{L}=s^{(0)}\cdot \frac{\rho^{(0)}}{1+\eta}.
\end{equation*}

\section{$\vev{\sigma^z_1\sigma_2^z}_\Psi$ from the Hellmann--Feynman theorem}

In this section we obtain an integral formula for the mean value
$\vev{\sigma^z_1\sigma_2^z}_\Psi$ for arbitrary excited states with a
smooth density of roots. 

Consider an eigenstate
\begin{equation*}
\ket{\Psi}=
\ket{\{\lambda\}_N}
\end{equation*}
in a finite spin chain of length $L$.
We apply the Hellmann--Feynman theorem \cite{Hellmann-Feynman} for this particular state, such
that we choose $\Delta$ as the variation parameter:
\begin{equation*}
  \bra{\Psi} \partial H/\partial \Delta
  \ket{\Psi}=
L\bra{\Psi} (\sigma^z_1\sigma^z_2-1)
  \ket{\Psi}
=\frac{\partial E_\Psi}{\partial \Delta}.
\end{equation*}
It is convenient to express the energy in terms of the $Q_2$
eigenvalue as given by the relation \eqref{Q2}.
Using $\Delta=\cosh(\eta)$ we obtain
\begin{equation*}
  L\bra{\Psi} (\sigma^z_1\sigma^z_2-1)\ket{\Psi}
=
2\coth(\eta) \vev{Q_2}_\Psi+
2\frac{\partial \vev{Q_2}_\Psi}{\partial \eta}.
\end{equation*}
In the following we evaluate the derivative on the r.h.s. in the
finite volume case and 
take the thermodynamic limit afterwards.

\subsection{States consisting of one-strings only}

First we consider the cases when the state $\ket{\Psi}$ consists of
1-strings only. 
Taking the derivative of the logarithm of the Bethe equations \eqref{BAe}
with respect to $\eta$ leads to 
\begin{equation}
\label{ij}
\begin{split}
0=  L  \tilde s^{(0)}(\lambda_j)+
L f(\lambda_j)s^{(0)}(\lambda_j)
+\sum_{k \ne j} \tilde\varphi(\lambda_j-\lambda_k)
+
\sum_{k \ne j}  (f(\lambda_j)-f(\lambda_k))
\varphi(\lambda_j-\lambda_k),
\end{split}
\end{equation}
where we introduced the ''shift function'' $f(\lambda)$ by
\begin{equation*}
  \frac{\partial \lambda_j}{\partial \eta}=f(\lambda_j)
\end{equation*}
and 
\begin{equation}
\label{q2most2}
\begin{split}
  \tilde s^{(0)}(u)&=(-i)\frac{\partial}{\partial \eta}  \log P(u)=\\
&=\frac{1}{2} \left[
\cot(u+i\eta/2)+\cot(u-i\eta/2)
\right]=
\frac{\sin(2u)}{2\sin(u-i\eta/2)\sin(u+i\eta/2)}\\
 \tilde \varphi(u)&=(-i)\frac{\partial}{\partial \eta}  \log S(u)=\\
&=- \left[
\cot(u+i\eta)+\cot(u-i\eta)
\right]=
\frac{-\sin(2u)}{\sin(u-i\eta)\sin(u+i\eta)}.
\end{split}
\end{equation}
In the thermodynamic limit
\eqref{ij}  is written as
\begin{equation}
\label{ijjk}
\begin{split}
0=   \tilde s^{(0)}(\lambda)+
 f(\lambda)s^{(0)}(\lambda)
+
\int\frac{d\omega}{2\pi} \rho^{(0)}_r(\omega) \tilde\varphi(\lambda-\omega)
+
\int\frac{d\omega}{2\pi} \rho^{(0)}_r(\omega) (f(\lambda)-f(\omega)) \varphi(\lambda-\omega).
\end{split}
\end{equation}
In the 1-string case the densities satisfy
\begin{equation}
\label{rhoegyenlet1s}
  \rho^{(0)}(u)=-s^{(0)}(u)-\int \frac{d\omega}{2\pi} \varphi(u-\omega) \rho^{(0)}_r(\omega)
\end{equation}
with $\varphi(u)=\varphi_{11}(u)$.
Substituting \eqref{rhoegyenlet1s} to \eqref{ijjk} leads to
\begin{equation}
\label{fegyenlet}
\begin{split}
0=   \tilde s^{(0)}(\lambda)-
 f(\lambda)\rho^{(0)}(\lambda)
+
\int\frac{d\omega}{2\pi} \rho^{(0)}_r(\omega) \tilde\varphi(\lambda-\omega)
-
\int\frac{d\omega}{2\pi} \rho^{(0)}_r(\omega) f(\omega) \varphi(\lambda-\omega).
\end{split}
\end{equation}
This equation uniquely determines $f(\lambda)$.
The variation of $Q_2$ is then expressed as
\begin{equation}
\label{kozt}
  \frac{\partial \vev{Q_2}_\Psi}{\partial \eta}=\sum 
\tilde s^{(1)}(\lambda_j)+\sum s^{(1)}(\lambda_j) f(\lambda_j),
\end{equation}
where
\begin{equation*}
\begin{split}
 \tilde s^{(1)}(u)&=\frac{\partial s^{(0)}(u)}{\partial \eta}=
-\frac{1}{2} \left[\frac{1}{\sin^2(u+i\eta/2)}+\frac{1}{\sin^2(u-i\eta/2)}\right]\\
 s^{(1)}(u)&=\frac{\partial s^{(0)}(u)}{\partial u}=
i
\left[\frac{1}{\sin^2(u+i\eta/2)}-\frac{1}{\sin^2(u-i\eta/2)}\right].
\end{split}
\end{equation*}
In the thermodynamic limit \eqref{kozt} is expressed as
\begin{equation}
\label{koz}
\frac{1}{L}  \frac{\partial \vev{Q_2}_\Psi}{\partial \eta}=\int \frac{du}{2\pi}
\big(\tilde s^{(1)}(u)\rho^{(0)}_r(u)+s^{(1)}(u)f(u)\rho^{(0)}_r(u)\big).
\end{equation}
Introducing the function
\begin{equation*}
  \sigma^{(0)}(u)=f(u)\rho^{(0)}(u)
\end{equation*}
and using the definition \eqref{etak} for the 1-strings we have
\begin{equation}
\label{fegyenlet2}
\begin{split}
0=   \tilde s^{(0)}(\lambda)-
 \sigma^{(0)}(\lambda)
+
\int\frac{d\omega}{2\pi} \rho^{(0)}_r(\omega) \tilde\varphi(\lambda-\omega)
-
\int\frac{d\omega}{2\pi} \frac{\sigma^{(0)}(\omega)}{1+\eta(\omega)} \varphi(\lambda-\omega)
\end{split}
\end{equation}
and
\begin{equation}
\label{koz2}
\frac{1}{L}  \frac{\partial \vev{Q_2}_\Psi}{\partial \eta}=\int \frac{du}{2\pi}
\left(\tilde s^{(1)}(u)\frac{\rho^{(0)}(u)}{1+\eta(u)}+
s^{(1)}(u)\frac{\sigma^{(0)}(u)}{1+\eta(u)}\right).
\end{equation}
The expectation value of $Q_2$ is simply
\begin{equation*}
  \frac{\vev{Q_2}_\Psi}{L}= \int \frac{du}{2\pi} s^{(0)}(u) \rho^{(0)}_r(u).
\end{equation*}
Putting everything together the final result for the mean value can be
written as
\begin{equation}
\label{MV1}
  \vev{ \sigma^z_1\sigma^z_2}_\Psi
=
1-\coth(\eta) \Omega_{0,0}+
\Gamma_{0,1},
\end{equation}
where
\begin{equation*}
  \Omega_{0,0}=-2\int \frac{du}{2\pi} s^{(0)}(u) \frac{\rho^{(0)}(u)}{1+\eta(u)}
\end{equation*}
and
\begin{equation*}
 \Gamma_{0,1}= 
2\int \frac{du}{2\pi}
\left(\tilde s^{(1)}(u)\frac{\rho^{(0)}(u)}{1+\eta(u)}+
s^{(1)}(u)\frac{\sigma^{(0)}(u)}{1+\eta(u)}\right).
\end{equation*}

\subsection{Multiple strings}

It is straightforward to extend the previous calculation to states with an
arbitrary string content. Instead of repeating all the steps we merely
present the results.  

The mean value $\vev{ \sigma^z_1\sigma^z_2}_\Psi$ can be expressed by the same
formula as in the 1-string case: 
\begin{equation}
\label{MV2}
  \vev{ \sigma^z_1\sigma^z_2}_\Psi
=1-
\coth(\eta) \Omega_{0,0}+
\Gamma_{0,1}.
\end{equation}
However, in the general case all strings contribute and we have
\begin{equation} 
\label{OmGam}
\begin{split}
  \Omega_{0,0}&=-2  s^{(0)} \cdot \frac{\rho^{(0)}}{1+\eta}\\
 \Gamma_{0,1}&= 
2\left(\tilde s^{(1)}\cdot\frac{\rho^{(0)}}{1+\eta}+
s^{(1)}\cdot \frac{\sigma^{(0)}}{1+\eta}\right).
\end{split}
\end{equation}
Here $\rho^{(0)}$ are the total root densities which are solutions to \eqref{rhoegyenlet2b}
and $\sigma^{(0)}$ is an auxiliary function satisfying
\begin{equation}
\label{fegyenlet3B}
\begin{split}
\sigma^{(0)}=   \tilde s^{(0)}+
\tilde\varphi\star \frac{\rho^{(0)}}{1+\eta}
-\varphi\star \frac{\sigma^{(0)}}{1+\eta}.
\end{split}
\end{equation}
In the formulas above we applied the notations introduced at the end
of Section 2.
The sources and kernels
follow from the appropriate derivatives of the string propagators and
string-string scattering phases. The one-particle functions are
\begin{equation}
\label{smost}
\begin{split}
  s^{(0)}_{k}(u)&=
(-i) \left[
\cot(u+ik\eta/2)-\cot(u-ik\eta/2)
\right]\\
 s^{(1)}_k(u)&=
i
\left[\frac{1}{\sin^2(u+ik\eta/2)}-\frac{1}{\sin^2(u-ik\eta/2)}\right]\\
\tilde s^{(0)}_{k}(u)&=
\frac{k}{2} \left[
\cot(u+ik\eta/2)+\cot(u-ik\eta/2)
\right]\\
 \tilde s^{(1)}_k(u)&=
-\frac{k}{2} \left[\frac{1}{\sin^2(u+ik\eta/2)}+\frac{1}{\sin^2(u-ik\eta/2)}\right].
\end{split}
\end{equation}
The kernels $\varphi_{nm}$ are given by \eqref{varphikeplet}, whereas $\tilde
\varphi_{nm}$ reads
\begin{equation*}
\tilde  \varphi_{nm}=
  \begin{cases}
-(\tilde s^{(0)}_{|n-m|}+2\tilde s^{(0)}_{|n-m|+2}+2\tilde s^{(0)}_{|n-m|+4}+\dots+2\tilde s^{(0)}_{n+m-2}+\tilde s^{(0)}_{n+m})    & \text{if}\quad  m\ne n\\
-(2\tilde s_2^{(0)}+2\tilde s_4^{(0)}+\dots+2\tilde s_{2n-2}^{(0)}+\tilde s_{2n}^{(0)})   & \text{if}\quad m=n.
  \end{cases}
\end{equation*}
Equations \eqref{MV2}-\eqref{OmGam} are a central result of this
 work. They are valid for arbitrary states with a smooth
distribution of string centers given that the string hypothesis
holds. 

For the sake of completeness we note that the x-x correlator
can be expressed using the relation $\vev{H}/L=-\sinh(\eta)\Omega_{0,0}$  as
\begin{equation*}
 \vev{ \sigma^x_1\sigma^x_2}_\Psi =\frac{1}{2\sinh\eta} \Omega_{0,0}
-\frac{\cosh\eta}{2} \Gamma_{0,1}.
\end{equation*}

\section{The finite temperature case}

In this section we consider finite temperature mean values of 
local correlators. 
They are defined as
\begin{equation*}
  \vev{\ordo}_T=\lim_{L\to\infty}\frac{\text{Tr}\ e^{-\beta H}\ordo }{Z(L)},
  \qquad Z(L)=\text{Tr}\
  e^{-\beta H}=e^{-L\beta f(\beta)},
\end{equation*}
where $\beta$ is the inverse temperature and $f(\beta)$ is the free energy density.

Traditionally there are two different methods in the Bethe Ansatz
literature to treat the finite temperature problem: the Quantum
Transfer Matrix (QTM) method \cite{kluemper-QTM,kluemper-review}
and the Thermodynamic Bethe Ansatz (TBA) \cite{Takahashi-book}.
Whereas they give the same result
for the free energy density \cite{TBA-QTM-Kluemper-Takahashi}, only
the QTM was known to provide the 
correlators as well. On the other hand, the TBA yields the finite
temperature root distributions, which can be used as an input to our
equations \eqref{MV2}-\eqref{OmGam}. This way we obtain an
independent formula for $\vev{\sigma_1^z\sigma_2^z}_T$ which can be
compared to the known result of the QTM method.

In the following  we collect the relevant results of the
two approaches and we compare the predictions for the nearest
neighbour correlator.

\subsection{Short distance correlators in the QTM approach}

In the QTM approach the central object is the auxiliary function
$\fa(\lambda)$ which is a complex valued function over the complex
plain  and it is the solution of the
integral equation
\begin{equation}
\label{NLIE1}
\begin{split}
\log \fa(\lambda)=&2\beta h
-2\sinh(\eta)\beta 
q^{(0)}_+(\lambda)
-  \int_C \frac{d\omega}{2\pi i} K(\lambda-\omega) \log(1+\fa(\omega)).
\end{split}
\end{equation}
Here $\beta=1/T$ is the inverse temperature, 
\begin{equation*}
  K(u)=\frac{\sinh2\eta }{\sinh(u+\eta)\sinh(u-\eta)},
\end{equation*}
and
\begin{equation*}
 q^{(0)}_+(\lambda)=
\frac{\cosh(\lambda)}{\sinh(\lambda)} -
\frac{\cosh(\lambda+\eta)}{\sinh(\lambda+\eta)}.
\end{equation*}
The free energy is computed as
\begin{equation*}
  f=h-T  \int_C \frac{d\omega}{2\pi i} q^{(0)}_+(\omega) \log(1+\fa(\omega)).
\end{equation*}
The complex integrals run over a contour $C$ which depends on the
anisotropy $\Delta$. For $\Delta>1$ it can be chosen 
 as the union of two straight line segments:
\begin{equation*}
C= [i\pi/2-\alpha,-i\pi/2-\alpha]\cup
 [-i\pi/2+\alpha,i\pi/2+\alpha],
\end{equation*}
where $\alpha<\eta/2$ is an arbitrary parameter which has to be chosen
large enough so that the contour encircles all zeroes of the function
$1+\fa(\omega)$ in the domain $|\text{Im}(\omega)|\le\pi/2$, $|\text{Re}(\omega)|\le\eta/2$.

Multiple integral formulas for the finite temperature correlation
functions were first derived in \cite{QTM1}. Later it was shown in
\cite{XXZ-finite-T-factorization} that the multiple integrals can be
factorized, i.e. they can be expressed as sums of products of simple
integrals. The paper
\cite{XXZ-massive-corr-numerics-Goehmann-Kluemper} also presents
numerical data for the short range correlators for the 
$\Delta>1$ regime.

In the following
we present the formulas extracted from \cite{XXZ-finite-T-factorization}
which are necessary for the
practical computations.
Let us define the
 auxiliary functions
$G_a(\lambda)$ and $\tilde G_a(\lambda)$, $a=0\dots \infty$,
which are solutions to the linear integral equations
\begin{equation}
\label{NLIEmderiv}
\begin{split}
G_{j}(\lambda)=&
 q_-^{(a)}(\lambda)
+ \int_C \frac{d\omega}{2\pi i} 
K(\lambda-\omega)
\frac{G_a(\omega)}{1+\fa(\omega)}
\end{split}
\end{equation}
and
\begin{equation*}
 \tilde G_a(\lambda,\mu)=
\tilde q_-^{(a)}
+
\int_C \frac{d\omega}{2\pi i}
\frac{G_a(\omega)}{1+\fa(\omega)} 
\tilde K(\lambda-\omega) +
\int_C \frac{d\omega}{2\pi i}
\frac{\tilde G_a(\omega)}{1+\fa(\omega)} 
K(\lambda-\omega),
\end{equation*}
where
\begin{equation*}
\begin{split}
  \tilde q_-^{(a)}&= \left(-\frac{\partial}{\partial\lambda}\right)^a 
\coth(\lambda-\eta)\\
  q_-^{(a)}&= \left(-\frac{\partial}{\partial\lambda}\right)^a
\left(\coth(\lambda-\eta)-\coth(\lambda)\right)
\end{split}
\end{equation*}
and 
\begin{equation*}
 \tilde  K(u)=\frac{\sinh(2u) }{\sinh(u+\eta)\sinh(u-\eta)}.
\end{equation*}
Let the numbers $\Psi_{a,b}$ and $P_{a,b}$ with $a,b=0\dots
\infty$ be given by
\begin{equation}
\label{PsiP}
\begin{split}
    \Psi_{a,b}&=
\int_C\frac{d\omega}{\pi i}
q_-^{(b)}(\omega)
\frac{G_a(\omega)}{1+\fa(\omega)}\\
  P_{a,b}&=
\int_C\frac{d\omega}{\pi i}\left[
q_-^{(b)}(\omega)
\frac{\tilde G_a(\omega)}{1+\fa(\omega)}
+
\tilde q_-^{(b)}(\omega)
\frac{G_a(\omega)}{1+\fa(\omega)}\right].
\end{split}
\end{equation}
These objects are the building blocks for the local correlations.
As a final step we define
\begin{equation}
\label{omW}
  \begin{split}
    \omega_{a,b}&=-\Psi_{a,b} -(-1)^b\frac{1}{2}
    \left(\frac{\partial}{\partial u}\right)^{a+b} K(u)\Big|_{u=0}\\
W_{a,b}&=
-P_{a,b} +(-1)^b\frac{1}{2}
    \left(\frac{\partial}{\partial u}\right)^{a+b} \tilde K(u)\Big|_{u=0}.
  \end{split}
\end{equation}
They possess the symmetry properties
\begin{equation*}
  \omega_{a,b}=\omega_{b,a}\qquad
\qquad
W_{a,b}=-W_{b,a}.
\end{equation*}
All short distance correlators can be expressed as a combination of
the numbers $\omega_{a,b}$ and $W_{a,b}$. The formulas can be found in
the papers
\cite{XXZ-finite-T-factorization,XXZ-massive-corr-numerics-Goehmann-Kluemper}
\footnote{Our notation differs slightly from 
\cite{XXZ-finite-T-factorization,XXZ-massive-corr-numerics-Goehmann-Kluemper}:
Our $\omega$ and $W$ correspond to $\omega$ and $\omega'/\eta$ of \cite{XXZ-finite-T-factorization,XXZ-massive-corr-numerics-Goehmann-Kluemper} ,
respectively, and we denote the number of derivatives with respect to
the variables $x$ and $y$ by $a$ and $b$. For example
$W_{1,2}=\omega'_{xyy}/\eta$. 
}.
Here we present four examples:
\begin{equation}
\label{corrpeldak}
  \begin{split}
\vev{\sigma^z_1\sigma^z_2}_T&=\coth(\eta)\omega_{0,0}+W_{1,0}\\
\vev{\sigma^x_1\sigma^x_2}_T&=-\frac{\omega_{0,0}}{2\sinh(\eta)}-\frac{\cosh(\eta)}{2} W_{1,0}\\
\vev{\sigma^z_1\sigma^z_3}_T&=2\coth(2\eta)\omega_{0,0}+W_{1,0}+\tanh(\eta)\frac{\omega_{2,0}-2\omega_{1,1}}{4}
-\frac{\sinh^2(\eta)}{4}W_{2,1}\\
\vev{\sigma^x_1\sigma^x_3}_T&=-\frac{1}{\sinh(2\eta)}\omega_{0,0}-\frac{\cosh(2\eta)}{2}
W_{1,0}-\tanh(\eta)\cosh(2\eta)\frac{\omega_{2,0}-2\omega_{1,1}}{8}+\\
&\hspace{6cm} +\sinh^2(\eta)\frac{W_{2,1}}{8}.
  \end{split}
\end{equation}
Using \eqref{omW} the nearest neighbour $z-z$ correlator is expressed as
\begin{equation}
\label{MV3}
  \vev{\sigma^z_1\sigma^z_2}_T=1-\coth(\eta)\Psi_{0,0}+P_{0,1},
\end{equation}
where we used $P_{0,1}=-P_{1,0}$.

\subsection{The nearest neighbour correlator from TBA}

The Thermodynamic Bethe Ansatz was devised to determine the free
energy density 
by finding a saddle point distribution to the free energy functional
\cite{YangYang2,Takahashi-book}. In the case of the XXZ spin chain
with $\Delta>1$ the TBA yields the following set of nonlinear integral
equations for the functions $\eta_k(u)$ defined in \eqref{etak}:
\begin{equation}
  \label{TBA}
  \log \eta_k(u)=-2k \beta h+2\sinh(\eta)\beta s^{(0)}_k(u) +
\sum_{j=1}^\infty\int_{-\pi/2}^{\pi/2} \frac{d\omega}{2\pi}
\varphi_{kj}(u-\omega)
\log(1+1/\eta_j(\omega)).
\end{equation}
The free energy density can then be expressed as
\begin{equation*}
f=h+T \sum_{k=1}^\infty 
\int_{-\pi/2}^{\pi/2} \frac{du}{2\pi}  s^{(0)}_k(u) \log(1+1/\eta_j(\omega)).
\end{equation*}
In \eqref{TBA} all $\eta_k$ are coupled through the kernels
$\varphi_{kj}$. A simpler form of the TBA equations can be derived where only
neighbouring equations are coupled \cite{Takahashi-book}; however, it
will not be used here.

If the $\eta_k$ are found from \eqref{TBA} then 
formulas
\eqref{MV2}-\eqref{fegyenlet3B}  can be used to find the
 densities, the auxiliary functions $\sigma^{(0)}_k$ and finally the finite
temperature mean value $\vev{\sigma_1^z\sigma_2^z}_T$. 

In order to check our results we numerically computed $\vev{\sigma_1^z\sigma_2^z}_T$ for different
values of $\Delta$, $\beta$ and $h$ from both the TBA and the QTM
formulas. We found convincing numerical evidence 
that the two approaches give the same results; examples of the numerical data are presented
in the Appendix.

\section{Conjectures for the short range correlators}

It is hard to miss the conspicuous similarities between
the QTM
and Bethe Ansatz (BA) formulas for $\vev{\sigma_1^z\sigma_2^z}_T$: there is a formal one-to-one
correspondence at each step of the calculation, even though on the BA
side we have an infinite number of equations, whereas on the QTM side
there are only two. It is easy to see that the building blocks of the formulas
coincide:
\begin{equation}
\label{iden1}
  \Omega_{0,0}=\Psi_{0,0}\qquad\qquad
\Gamma_{0,1}=P_{0,1}.
\end{equation}
The first equation follows from the fact that both $\Omega_{0,0}$ and
$\Psi_{0,0}$ are proportional to the mean value of the
Hamiltonian, whereas the second follows from  the equality of the two
formulas for the correlator given that the first relation is already established.

Based on these observations we generalize the formulas
 of Section 2 as follows.
Let  $\rho^{(a)}$ and $\sigma^{(a)}$ for general $a\ge 0$ be given as
the unique solution to the linear equations
\begin{equation}
\label{rhoegyenlet2C}
  \rho^{(a)}=-s^{(a)}-
\varphi \star \frac{\rho^{(a)}}{1+\eta}
\end{equation}
\begin{equation}
\label{fegyenlet3C}
\begin{split}
\sigma^{(a)}=   \tilde s^{(a)}+
\tilde\varphi\star \frac{\rho^{(a)}}{1+\eta}
-\varphi\star \frac{\sigma^{(a)}}{1+\eta}.
\end{split}
\end{equation}
Here 
\begin{equation*}
  s^{(a)}(u)=\left(\frac{\partial}{\partial u}\right)^a s^{(0)}(u)\qquad\qquad
\tilde  s^{(a)}(u)=\left(\frac{\partial}{\partial u}\right)^a \tilde s^{(0)}(u).
\end{equation*}
We define the numbers $\Omega_{a,b}$ and $\Gamma_{a,b}$ as
\begin{equation} 
\label{Omjk}
  \Omega_{a,b}=-2  s^{(b)} \cdot \frac{\rho^{(a)}}{1+\eta}
\end{equation}
and
\begin{equation}
\label{Gjk}
 \Gamma_{a,b}= 
2\left(\tilde s^{(b)}\cdot\frac{\rho^{(a)}}{1+\eta}+
s^{(b)}\cdot \frac{\sigma^{(a)}}{1+\eta}\right).
\end{equation}

We propose the following two conjectures.

\begin{conj}
In the finite temperature situation the building blocks of the
QTM formulas for correlators can be
computed from the string densities only. That is,
\begin{equation}
\label{azonos}
  \Omega_{a,b}=(-1)^{(a+b)/2}\Psi_{a,b} \qquad\text{and}\qquad
\Gamma_{a,b}=(-1)^{(a+b-1)/2}P_{a,b},\qquad a,b=0\dots \infty,
\end{equation}
given that $\Omega_{a,b}$ and $\Gamma_{a,b}$ are calculated from
\eqref{rhoegyenlet2C}-\eqref{Gjk} with $\eta_k(u)$ being the solution
of the TBA equation 
\eqref{TBA} and $\Psi_{a,b}$ and $P_{a,b}$ are calculated from \eqref{PsiP} with
$\fa(u)$ being the solution of the NLIE \eqref{NLIE1} with the same
inverse temperature and magnetic field.
\end{conj}

The sign factors in \eqref{azonos} can be
understood by noticing that the QTM and TBA rapidity parameters are
related by a rotation of $90^\circ$ in the complex plane. This results
in factors of $i$ whenever the formulas for the sources are differentiated.
The overall sign of $\Omega$ and $\Gamma$ is fixed by the
identities \eqref{iden1} which follow from their definition and the
Hellmann--Feynman theorem.

We numerically checked Conjecture 1 and convinced ourselves that it holds for
arbitrary anisotropy, temperatures and magnetic fields; examples of
the correlators computed using Conjecture 1 are presented in the Appendix.
It follows  that the finite temperature
correlators can be obtained from the TBA string densities only.
While this procedure is numerically more costly than the pure
QTM calculation, our second conjecture states that the validity of the
Bethe Ansatz formulas might be 
more general and not restricted to the finite temperature situation. 

\begin{conj}
Short distance correlators in any Bethe state with smooth
distribution of string centers can be computed via the following
procedure:
\begin{enumerate}
\item Compute the auxiliary functions $\rho^{(a)}$ and $\sigma^{(a)}$
  from  \eqref{rhoegyenlet2C} and \eqref{fegyenlet3C}.
\item Compute $\Omega_{a,b}$ and $\Gamma_{a,b}$ from
  \eqref{Omjk} and \eqref{Gjk}.
\item Use the relations
\begin{equation}
\label{omWW}
  \begin{split}
    \omega_{a,b}&=-(-1)^{(a+b)/2}\Omega_{a,b} -(-1)^b\frac{1}{2}
    \left(\frac{\partial}{\partial u}\right)^{a+b} K(u)\Big|_{u=0}\\
W_{a,b}&=
-(-1)^{(a+b-1)/2}\Gamma_{a,b} +(-1)^b\frac{1}{2}
    \left(\frac{\partial}{\partial u}\right)^{a+b} \tilde K(u)\Big|_{u=0}
  \end{split}
\end{equation}
 to obtain $\omega_{a,b}$ and
  $W_{a,b}$.
\item Substitute $\omega_{a,b}$ and  $W_{a,b}$ into the already available
factorized formulas of the QTM literature. Examples for two-site and
three-site correlators are presented in \eqref{corrpeldak}.
\end{enumerate}
\end{conj}

We have shown that Conjecture 2 is true for the nearest neighbour
correlator with arbitrary string distributions, and numerical evidence
supports its validity for all local correlators in the finite
temperature case. It is an open question whether it is true in full generality.

\section{Discussion and Outlook}

We derived an analytical formula for the simplest non-trivial
correlator $\vev{\sigma_1^z\sigma_2^z}_\Psi$ which is valid for Bethe
states with arbitrary string distributions. Our result could be used
in quantum quench problems to give prediction for the long-time limit
of this particular observable, if the root densities can be determined
from the quench action \cite{quench-action,caux-talk,JS-oTBA,sajat-oTBA}. 

Based on very close similarities
with known formulas from the QTM approach
we conjectured a generalization to other short distance
correlators. If Conjecture 2 is found to be true, it would give a
new interpretation to the factorized formulas of
\cite{XXZ-finite-T-factorization}. And even if it is not true, it
still needs to be understood why it works in the finite temperature
case (see Conjecture 1 and the numerical data in the Appendix).

We think it is worthwhile to recall a situation which seems to
be analogous to the present one. In 
\cite{LM-sajat} an infinite integral series was derived for the $K$-body
 local correlators of the 1D Bose gas; the idea for this series originated from the theory
of integrable QFT's where it is called the LeClair--Mussardo (LM) series. The
 result of  \cite{LM-sajat} holds for an arbitrary distribution of
 Bethe roots. In \cite{g3-exact} the LM series was summed up into 
 simple formulas in the cases $K=2,3$. Later a different
 multiple integral formula was found in \cite{XXZ-to-LL-sajat} which
 was factorized for $K\le 4$, reproducing the earlier results of
 \cite{g3-exact}. It is very important that all of these calculations
 are completely general in the sense that they do not rely on any assumption of an
 underlying thermal ensemble. The only input to the factorized
 formulas is the filling fraction
 $\rho_{r}(\lambda)/\rho(\lambda)$, which is treated as an arbitrary function.

We believe that a LeClair--Mussardo series could be established 
also for the local operators of the XXZ chain using the methods of
\cite{LM-sajat}. If such a series exists, then its application for ground
state or finite temperature correlations would be far less effective than any
of the existing methods, but it would be valid for arbitrary
string distributions. Moreover, the conjectured results of the
present work could follow from a summation of the LM series. We leave
these questions to further research.

\vspace{1cm}
{\bf Acknowledgments} 

\bigskip

We are grateful to M\'arton Kormos and G\'abor Tak\'acs for inspiring
discussions and for useful comments on the manuscript.

\bigskip
\bigskip

{\bf Notes added:} 

The conjectured formulas of this work were used in the
paper \cite{sajat-oTBA} to give predictions for the long-time limit of
local correlations following a quantum quench. In all cases
perfect agreement was found with the results of real-time numerical
simulations, and this provides strong evidence for the validity of
Conjecture 2. 

Our results for the nearest neighbour z-z correlator presented in
Section 3 were obtained independently in the paper \cite{JS-oTBA},
which was submitted as an e-print to the arXiv on the same day as
the present work.

\appendix

\section{Numerical checks}

We numerically checked Conjecture 2 in the finite temperature
situation by computing both the QTM and TBA formulas for the
quantities $\omega_{a,b}$ and $W_{a,b}$ and substituting them into the
factorized formulas for the short distance two-point
functions. We truncated the infinite set of TBA equations and observed
that the truncation number needed to obtain good agreement with the
QTM results strongly depends on the temperature and the magnetic
field. For large negative magnetic fields and small temperatures the higher strings
are suppressed; we observed that in many cases perfect numerical agreement can be
found already with $\le 10$ equations. On the other hand, if at a
finite (non-zero) temperature the magnetic field is zero or very small 
then the contributions of the higher strings can be considerable and a
large number of equations is required. In such cases we found that
instead of equations \eqref{rhoegyenlet2C}-\eqref{fegyenlet3C} it
is more efficient to use the decoupled form which can be written as follows.

Consider an equation of the type
\begin{equation}
\label{integral12}
  f_k(u)=g_k(u)-\sum_{l=1}^\infty \int_{-\pi/2}^{\pi/2} 
\frac{d\omega}{2\pi} \varphi_{kl}(u-\omega) 
\frac{f_k(\omega)}{1+\eta_k(\omega)}.
\end{equation}
Applying the well-known steps to decouple the convolution  terms
 results in \cite{Takahashi-book}
\begin{equation}
\label{integral12b}
\begin{split}
&  f_k(u)=g_k(u)-\int_{-\pi/2}^{\pi/2} 
\frac{d\omega}{2\pi} s(u-\omega) (g_{k-1}(u)+g_{k+1}(u))+\\
&\hspace{3cm}+ \int_{-\pi/2}^{\pi/2} 
\frac{d\omega}{2\pi} s(u-\omega) 
\left(\frac{\eta_{k+1}(\omega)f_{k+1}(\omega)}{1+\eta_{k+1}(\omega)}+
\frac{\eta_{k-1}(\omega)f_{k-1}(\omega)}{1+\eta_{k-1}(\omega)}\right),
\end{split}
\end{equation}
where
\begin{equation*}
  s(x)=
{1}+2\sum_{n=1}^\infty \frac{\cos(2n x)}{\cosh(\eta n)}.
\end{equation*}
This means a drastic simplification in the equation 
 of $\rho^{(a)}$ because all source terms disappear except for $k=1$:
\begin{equation}
\label{integral12bc}
\begin{split}
&  \rho^{(a)}_k(u)=\delta_{k,1}\left(\frac{\partial}{\partial u}\right)^as(u)+\\
&\hspace{3cm}+ \int_{-\pi/2}^{\pi/2} 
\frac{d\omega}{2\pi} s(u-\omega) 
\left(\frac{\eta_{k+1}(\omega)\rho^{(a)}_{k+1}(\omega)}{1+\eta_{k+1}(\omega)}+
\frac{\eta_{k-1}(\omega)\rho^{(a)}_{k-1}(\omega)}{1+\eta_{k-1}(\omega)}\right).
\end{split}
\end{equation}
There is a similar formula for $\sigma^{(a)}$ which will be presented elsewhere.

Examples of our numerical data are shown in Table 1. We computed the local
correlators $\vev{\sigma^a_1\sigma_{1+j}^a}_T$ with $a=z,x$ and
$j=1,2,3$ for various values of the triplets $(\Delta,\beta,h)$. For
the sake of brevity here we only present results for 
$\vev{\sigma^z_1\sigma_{2}^z}_T$ and
$\vev{\sigma^x_1\sigma_{4}^x}_T$ in 4 different situations. 

In the first
case we chose a relatively large magnetic field. Here $N=10$ equations
sufficed to obtain the correct result up to 8 digits; we used the
original linear equations. Note that even though the $N=10$ result is
already accurate to the digits given, the $N=2$ and $N=4$ truncations still show
small differences. 

In the other three cases the magnetic field was chosen to be smaller or
zero. Accordingly a larger number of equations was required; in
these cases we
used the decoupled form of the linear integral equations. 
The TBA data always
seems to converge to the QTM result, but for very small
or zero magnetic fields we found a relative error of
the order $10^{-3}$ even at $N=40$, where our current computer programs are
already slow. We conclude that in these
situations the strings that are longer than 40 still contribute
considerably and our programs need to be improved to be able to handle
these cases as well.

\begin{table}
  \centering
\begin{subtable}{1.0\textwidth}
 \centering
  \begin{tabular}{|c|c|c|c||c|}
\hline
 &    $N=2$ & $N=4$ & $N=10$ & QTM  \\
\hline
$\vev{\sigma_1^z\sigma_2^z}_T$ & -1.7663637$\cdot 10^{-1}$ & -1.7678391 
$\cdot 10^{-1}$ & -1.7678511 $\cdot 10^{-1}$&  -1.7678511$\cdot 10^{-1}$\\
\hline
$\vev{\sigma_1^x\sigma_4^x}_T$ & -5.7948070$\cdot 10^{-3}$ &
-5.5484712$\cdot 10^{-3}$ & -5.5455688$\cdot 10^{-3}$ &-5.5455688
$\cdot 10^{-3}$ \\ 
\hline
  \end{tabular}
\caption{$\Delta=3$, $\beta=0.2$, $h=-1$}
\end{subtable}

\bigskip

\begin{subtable}{1.0\textwidth}
 \centering
  \begin{tabular}{|c|c|c|c||c|}
\hline
 &    $N=10$ & $N=20$ & $N=30$ & QTM  \\
\hline
$\vev{\sigma_1^z\sigma_2^z}_T$ & -4.7190677 $\cdot 10^{-1}$ & 
-4.7183624$\cdot 10^{-1}$ &  -4.7183623 $\cdot 10^{-1}$&-4.7183623  $\cdot 10^{-1}$\\
\hline
$\vev{\sigma_1^x\sigma_4^x}_T$ & -7.1256853$\cdot 10^{-3}$ &
 -6.9050701$\cdot 10^{-3}$ &  -6.9050442 $\cdot 10^{-3}$ &
-6.9050445$\cdot 10^{-3}$ \\ 
\hline
  \end{tabular}
\caption{$\Delta=3$, $\beta=0.2$, $h=-0.4$}
\end{subtable}

\bigskip

\begin{subtable}{1.0\textwidth}
 \centering
  \begin{tabular}{|c|c|c|c||c|}
\hline
 &    $N=10$ & $N=20$ & $N=40$ & QTM  \\
\hline
$\vev{\sigma_1^z\sigma_2^z}_T$ & -5.2791241 $\cdot 10^{-1}$ &
 -5.2480411 $\cdot 10^{-1}$ & -5.2459375 $\cdot 10^{-1}$& -5.24591662 $\cdot 10^{-1}$\\
\hline
$\vev{\sigma_1^x\sigma_4^x}_T$ &-1.227053385$\cdot 10^{-2}$ &
 -7.4127763 $\cdot 10^{-3}$ &  -7.1443727 $\cdot 10^{-3}$ &
-7.14182791$\cdot 10^{-3}$ \\
\hline
  \end{tabular}
\caption{$\Delta=3$, $\beta=0.2$, $h=-0.1$}
\end{subtable}

\bigskip

\begin{subtable}{1.0\textwidth}
 \centering
  \begin{tabular}{|c|c|c|c||c|}
\hline
 &    $N=10$ & $N=20$ & $N=40$ & QTM  \\
\hline
$\vev{\sigma_1^z\sigma_2^z}_T$ &-6.6205394 $\cdot 10^{-1}$ &
 -6.5927266 $\cdot 10^{-1}$ & -6.5869615  $\cdot 10^{-1}$& -6.5852520 $\cdot 10^{-1}$\\
\hline
$\vev{\sigma_1^x\sigma_4^x}_T$ &-5.0765570 $\cdot 10^{-3}$ &
-4.7597568$\cdot 10^{-3}$ & -4.7526870$\cdot 10^{-3}$ &
 -4.74751256$\cdot 10^{-3}$ \\
\hline
  \end{tabular}
\caption{$\Delta=2$, $\beta=0.5$, $h=0$}
\end{subtable}

  \caption{
Examples of the numerical results for the finite temperature local
correlators. The first three columns present 
    the calculations with the truncated TBA system with $N$ equations, whereas the last
    column is the numerically exact result of the QTM method. 
     }
  \label{tab:1}
\end{table}

\clearpage

\addcontentsline{toc}{section}{References}
\bibliography{../../pozsi-general.bib}
\bibliographystyle{utphys}

\end{document}